# Plasmonics at Radio Frequencies

Igor I. Smolyaninov, Quirino Balzano, John Mulholland, Viktor A. Podolskiy and Dendy Young

***Abstract*—We demonstrate that low loss radio frequency plasmon-like surface electromagnetic waves may propagate along rough interfaces between air and highly conductive media, such as water or metal. Unlike the well-known Zenneck wave solutions, these surface waves have mode index larger than 1, so that their propagation properties are generally similar to the properties of surface plasmons in the visible frequency range. The wave impedance of these modes is derived, which demonstrates the need to use special plasmonic inducers or exciters for efficient generation of these novel surface electromagnetic modes. Propagation characteristics of these "radio frequency plasmons" were explored both experimentally and theoretically in the cases of water-air and water-ice-air interfaces, and the two-dimensional character of this propagation has been revealed. Other important applications of radio frequency plasmons were demonstrated, such as communication along metal infrastructure immersed in water, communication along a lake floor, underwater plasmonic radar, as well as plasmonic communications underground and in the dense jungles.**

## I. Introduction

THE SURFACE electromagnetic wave solutions of Maxwell equations in the radio frequency range remained a somewhat controversial issue for more than hundred years [1]. One of the reasons for this controversy may be identified as treatment of an interface between two neighboring media as infinitely sharp, which inevitably leads to a surface wave dispersion law of the form

$$k_{sw} = \frac{\omega}{c}\left[\frac{\varepsilon_1\varepsilon_2}{\varepsilon_1+\varepsilon_2}\right]^{1/2} \qquad (1)$$

where $k_{sw}$ is the wavevector of the surface wave, ω is the frequency, and $\varepsilon_1$ and $\varepsilon_2$ are the frequency-dependent dielectric permittivities of the neighboring media [2]. Eq.(1) works very well in the visible frequency range, where it correctly predicts the dispersion law of surface plasmons on the surfaces of such low-loss metals as gold and silver, which have very large and negative real parts of their dielectric permittivity $\varepsilon_2$' and much smaller imaginary part $\varepsilon_2$". As may be seen from Eq.(1), it results in $k_{sw}>\omega/c$ for good metals if $\varepsilon_1=1$ is assumed, so that the resulting surface plasmon solutions are low-loss non-radiating surface waves, which are protected from radiative losses by momentum conservation [2].

Unfortunately, the same equation does not produce well-behaved surface electromagnetic wave solutions in the radio frequency range. At the low frequencies of radio waves the dielectric response of such media as typical metals and water is dominated by their electric conductivity, so that $\varepsilon_2$">>$\varepsilon_2$' and Eq.(1) results in the so called Zenneck wave, which has $k_{sw}<\omega/c$ [3]. Zenneck waves at the metal-air and water-air interfaces are not protected by momentum conservation. Therefore, they were argued to be unphysical mathematical oddities [4], since upon excitation, they are supposed to immediately decay into free space radio waves.

The next very recent development on this subject was reported in [5], where it was argued that the radiative character of Zenneck waves changes, and they become true non-radiative plasmon-like surface waves if one is willing to get rid of the unphysical treatment of interfaces as infinitely sharp. The natural roughness of metal surfaces and wavy water was shown to be sufficient for such a transition. In the next section we will confirm this important conclusion via perturbation theory and through numerical modeling of rough interfaces using COMSOL Multiphysics. These calculations confirm existence of plasmon-like surface electromagnetic waves in the radio frequency range, and they lead to theoretical prediction of numerous other real-world scenarios in which such "radio frequency plasmons" may be used to enhance wireless communications and sensing. These results open up a very broad range of new impactful applications of radio frequency plasmonics, which may potentially rival the plasmonic revolution in classical optics which happened at the turn of the current century [6,7].

## II. Tracing the Transition from Zenneck Waves to Radio Frequency Plasmons

The approach to surface wave problem developed in [5] consisted in replacing a sharp step-like transition between the dielectric permittivities $\varepsilon_1$ and $\varepsilon_2$ of the two neighboring media with some smooth continuous function ε(z), which equals $\varepsilon_2$ at

I. I. Smolyaninov is with Saltenna Inc., 1751 Pinnacle Dr, Ste 600 McLean, VA 22102-4007 USA (e-mail: igor.smolyaninov@ saltenna.com).

Q. Balzano is with Saltenna Inc., 1751 Pinnacle Dr, Ste 600 McLean, VA 22102-4007 USA (e-mail: quibalzano@gmail.com).

J. Mullholland is with Saltenna Inc., 1751 Pinnacle Dr, Ste 600 McLean, VA 22102-4007 USA (e-mail: john.mulholland@saltenna.com).

V. A. Podolskiy is with Saltenna Inc., 1751 Pinnacle Dr, Ste 600 McLean, VA 22102-4007 USA and Department of Physics and Applied Physics, University of Massachusetts Lowell, Lowell MA 01854 USA (e-mail:viktor.podolskiy@saltenna.com).

D. Young is with Saltenna Inc., 1751 Pinnacle Dr, Ste 600 McLean, VA 22102-4007 USA (e-mail:dendy.young@saltenna.com).

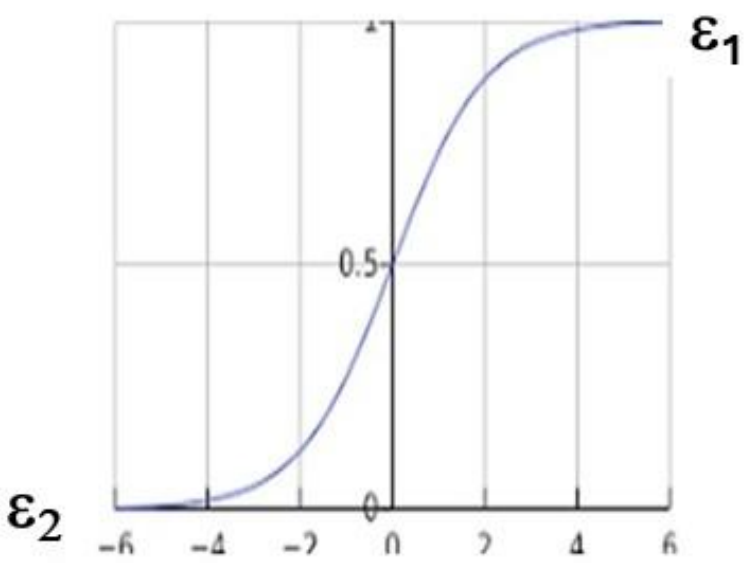

Fig. 1. Example of a sigmoid transition region between a medium having dielectric permittivity $\varepsilon_2$ and a medium having dielectric permittivity $\varepsilon_1$.

large negative z and $\varepsilon_1$ at large positive z. The resulting wave equation describing a TM polarized surface electromagnetic wave depends on ε(z) and its derivatives as follows [5]:

$$-\frac{\partial^2\psi}{\partial z^2}+\left(-\frac{\varepsilon(z)\omega^2}{c^2}-\frac{1}{2}\frac{\partial^2\varepsilon}{\varepsilon\partial z^2}+\frac{3}{4}\frac{(\partial\varepsilon/\partial z)^2}{\varepsilon^2}\right)\psi=-\frac{\partial^2\psi}{\partial z^2}+V\psi=-k^2\psi \quad (2)$$

where the effective wave function was introduced as $E_z=\psi/\varepsilon^{1/2}$, V(z) plays the role of effective potential, and the square of the surface wave wavevector taken with the minus sign $-k^2$ plays the role of effective energy in the one-dimensional Schrodinger equation (2). This wave equation was solved numerically for several transition functions ε(z), and plasmon-like solutions having $k>\omega/c$ were found in a large number of important cases [5]. However, no detailed analysis of such solutions existence has been performed.

In this work we perform a systematic analysis of surface wave characteristics for the particular case of a sigmoid distribution of ε(z) that provides a convenient way to analyze the transition between sharp discontinuity of permittivity and a more realistic, smooth interface, as depicted in Fig.1:

$$\varepsilon(z)=\frac{\varepsilon_1+\varepsilon_2 e^{\beta z}}{1+e^{\beta z}} \quad (3)$$

The sharpness parameter β in Eq.(3), controls the properties of the interface: the "conventional" step-discontinuous permittivity corresponds to (β→∞), while the realistic "smooth" boundary between two media corresponds to (β≠∞).

To analyze the modes supported by the smooth interface between two materials we reformulate Maxwell equations (or, equivalently, Eq(2)) in terms of the out-of-plane magnetic field, following the approach described in [8]:

$$\frac{\partial}{\partial z}\left(\frac{1}{\varepsilon}\frac{\partial H_\perp}{\partial z}\right)+\left(\frac{\omega^2}{c^2}-\frac{k_x^2}{\varepsilon}\right)H_\perp=0 \quad (4)$$

and solve this equation numerically with commercial finite-element-method solver (COMSOL Multiphysics [9]) and analytically, as explained below.

The results of numerical solutions of Eq.(4) for fixed operating frequency of 30 MHz for water-air interface are shown in Fig.2, where we use the dimensional mode index $n=k_x c/\omega$ to describe the mode propagation along the interface. It is seen that the limit of the extremely sharp interface yields the well-known Zenneck wave solution. Notably, the effective index of this (leaky) wave is below the free-space light cone. However, as the interface becomes smoother, the effective index of the mode increases, with the mode eventually crossing

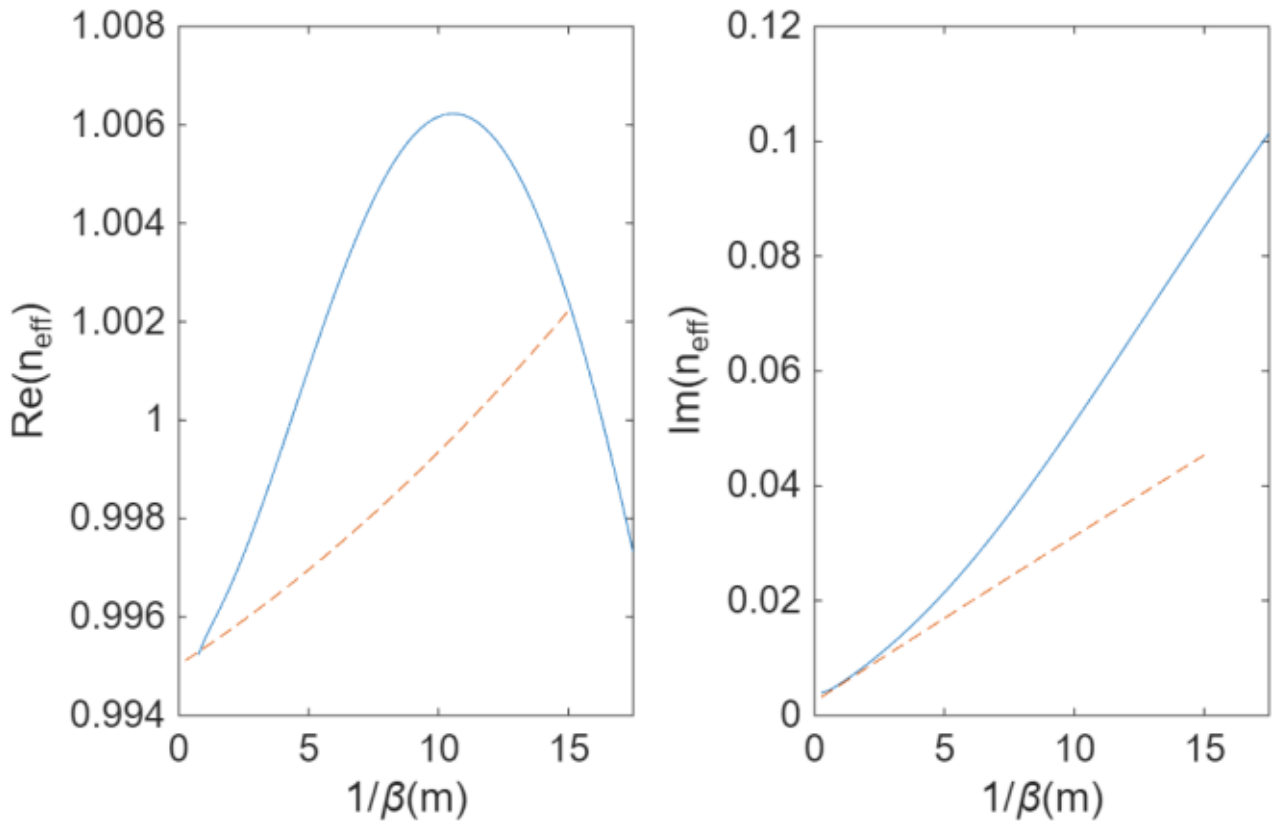

Fig. 2. Real (left panel) and imaginary (right panel) parts of the Zenneck-like mode index n as a function of interface sharpness parameter $\beta$, as calculated by full-wave numerical solutions of Maxwell equations with FEM (solid lines) and by first-order perturbation theory (dashed lines). Permittivity of water is set to 80+40i (fresh water), the operating frequency is set to 30 MHz

the light cone and exhibiting the guided mode characteristics, in agreement with predictions of Ref [5].

Alternatively, the change in the modal properties as a result of the change in the interface can be analyzed analytically using the first-order perturbation theory. Here, we assume that the field distribution and the propagation constants of the mode in real (smooth interface) system slightly deviate from the analytical solution given by the Zenneck wave. We then express the permittivity of the system, as well as the field distribution and propagation constant of the modes as those of the Zenneck wave $(\varepsilon_0(z), H_0(z), n_0)$ with (small) corrections $[\Delta\varepsilon(z), \Delta H(z), \Delta n]$, and linearize Eq.(4) in terms of these corrections. Finally, we use the orthogonality property of the waveguide modes $\int E_{z,m}^* H_{y,n} dz \propto \delta_{nm}$[10-12] (with $E^*$ being the field of the counter-propagating mode) to arrive at the following expression for $\Delta n$:

$$\Delta n=-\frac{\int \Delta\varepsilon\left(\frac{c^2}{\omega^2}\frac{dE_{z0}}{dz}\frac{dH_{y0}}{dz}+n_0^2\,E_0 H_0\right)dz}{2\,n_0\int \varepsilon_0\,E_0\,H_0\,dz} \quad (5)$$

Predictions of Eq.(5) are compared with numerical solutions of Maxwell equations in Fig.2. It is seen that the first-order perturbation theory reasonably describes the correction to the propagation constant of the mode in the limit $\frac{\beta c}{\omega}\ll 1$. In particular, the perturbation approach predicts the conversion of the leaky Zenneck wave into the guided mode. The higher-order corrections to propagation constant become important for larger values of the sharpness parameter (smoother interfaces), where the dependence of effective index on $\beta$ is strongly nonlinear.

The conversion of the leaky Zenneck wave into truly guided mode represents one of the main results of this work. Since no real interface is infinitely sharp and since the properties of realistic water-air interfaces are always affected by waves (often, with subwavelength wavelength), the mode supported by the water-air interface represents a true plasmon-like surface electromagnetic mode, which is protected from radiative losses by momentum conservation. We should note the low-loss long propagation length character of this "radio frequency plasmon". As we will discuss in the next sections of this paper, essentially similar results may be obtained in many other situations of great practical importance for wireless communications and sensing, such as the cases of metal, ground and ice surfaces, as well as the boundaries of foliage layers. These results highlight novel capabilities of radio frequency plasmonics, which may potentially replicate the recent plasmonics revolution in the optical domain [6,7].

Fig. 3. Radiation patterns of conventional and plasmonic antennas near conductive surfaces complement each other.

## III. Wave impedance of the radio frequency plasmons

While the exciting future perspectives of radio frequency plasmonics are clearly based on the recent history of its optical counterpart, it will be very important to develop efficient means of excitation and detection of radio frequency plasmons. Similar to plasmonic nanoantennas developed in the visible frequency range [13], we need to develop efficient exciters for plasmons at radio frequencies. In general, this task requires thorough understanding of the wave impedance of radio frequency plasmons.

To achieve this goal, let us further develop the mathematical formalism introduced in [5]. Let us consider an interface between two media characterized by the dielectric permittivity profile ε(z), and solve the corresponding Maxwell equations in cylindrical coordinates (assuming that a notional plasmonic exciter also exhibits cylindrical symmetry). The covariant macroscopic Maxwell equations written in cylindrical coordinates $(x^1,x^2,x^3)=(r,\phi,z)$ may be found in [14]. Since we are interested only in the special case of TM modes having zero angular momentum (m=0), the only nonzero field components in these modes are $E_1$, $E_3$ and $B_2$, resulting in the wave equation (6).

$$\frac{\partial^2 B_2}{\partial z^2}-\frac{(\partial\varepsilon/\partial z)}{\varepsilon}\frac{\partial B_2}{\partial z}+\frac{\partial^2 B_2}{\partial r^2}-\frac{1}{r}\frac{\partial B_2}{\partial r}+\frac{\omega^2\varepsilon}{c^2}B_2=0 \qquad (6)$$

By introducing an effective wave function as $B_2=\psi(r,z)\varepsilon^{1/2}r^{1/2}$, the wave equation (6) may be re-written and split as

$$-\frac{\partial^2\psi}{\partial z^2}+\left(-\frac{\varepsilon(z)\omega^2}{c^2}-\frac{1}{2}\frac{\partial^2\varepsilon}{\varepsilon\partial z^2}+\frac{3}{4}\frac{(\partial\varepsilon/\partial z)^2}{\varepsilon^2}\right)\psi=\frac{\partial^2\psi}{\partial r^2}-\frac{3}{4r^2}\psi=-k^2\psi \qquad (7)$$

If k is close to a real number (for a low loss plasmonic mode that we want to excite), solutions of Eq.(7) may be written as

$$\psi=\phi(z)r^{1/2}J_1(kr), \qquad (8)$$

where the z-dependent part satisfies Eq.(2), and the real physical field $B_\phi$ equals approximately

$$B_\phi^{real}=\phi(z)\varepsilon^{1/2}J_1(kr). \qquad (9)$$

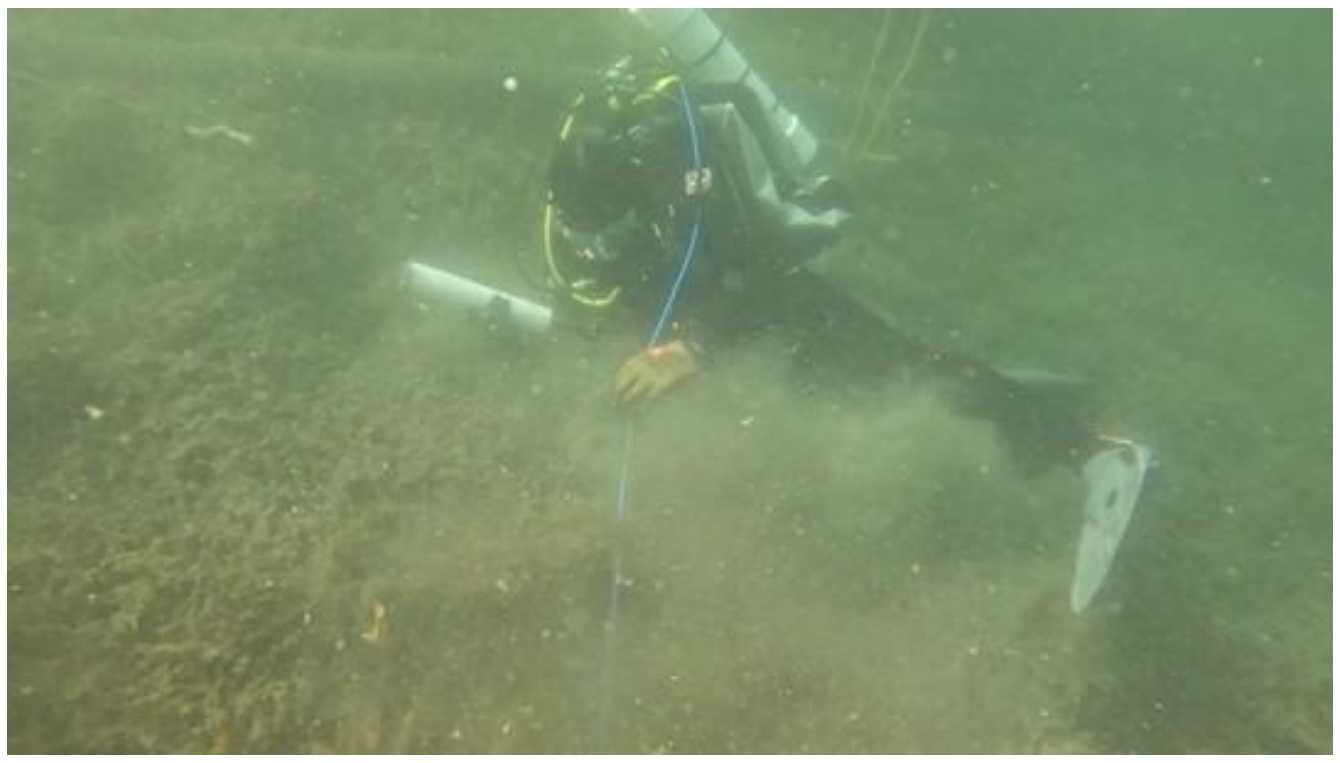

Fig. 4. A diver carrying a 30 MHz underwater plasmonic exciter in his hand.

if k is approximately real. As a result, the ratios of the real physical E and B field components may be written as

$$\frac{E_r}{B_\phi}=-\frac{ic}{\omega\varepsilon^{3/2}\phi(z)}\frac{\partial}{\partial z}\left(\varepsilon^{1/2}\phi(z)\right) \qquad (10)$$

and

$$\frac{E_z}{B_\phi}=-\frac{ic}{\omega\varepsilon rJ_1(kr)}\frac{\partial}{\partial r}\left(rJ_1(kr)\right)=\frac{ick}{\omega\varepsilon}\frac{J_0(kr)}{J_1(kr)} \qquad (11)$$

If k is approximately real, the radial component of the wave impedance behaves as an effective capacitance, while the z component of the wave impedance behaves as an effective inductance. Therefore, unlike ordinary free space antennas, the role of plasmonic antennas (or exciters) consist in providing the most efficient coupling from an input transmission line (a coaxial cable) to an effective transmission line at an interface between two media, which supports a propagating radio frequency surface electromagnetic mode, and which have very high reactance. While some particular plasmonic antenna designs operating near metal-air and water-air interfaces have been described in [15,16], the described formalism provides a useful recipe for a generic plasmonic antenna design goals.

We should also add that based on Eq.(6-8) the radiation patterns of conventional and plasmonic antennas near conductive surfaces (see Fig.3) complement each other, which indicates a very large number of interesting sensing applications of radio frequency plasmonics.

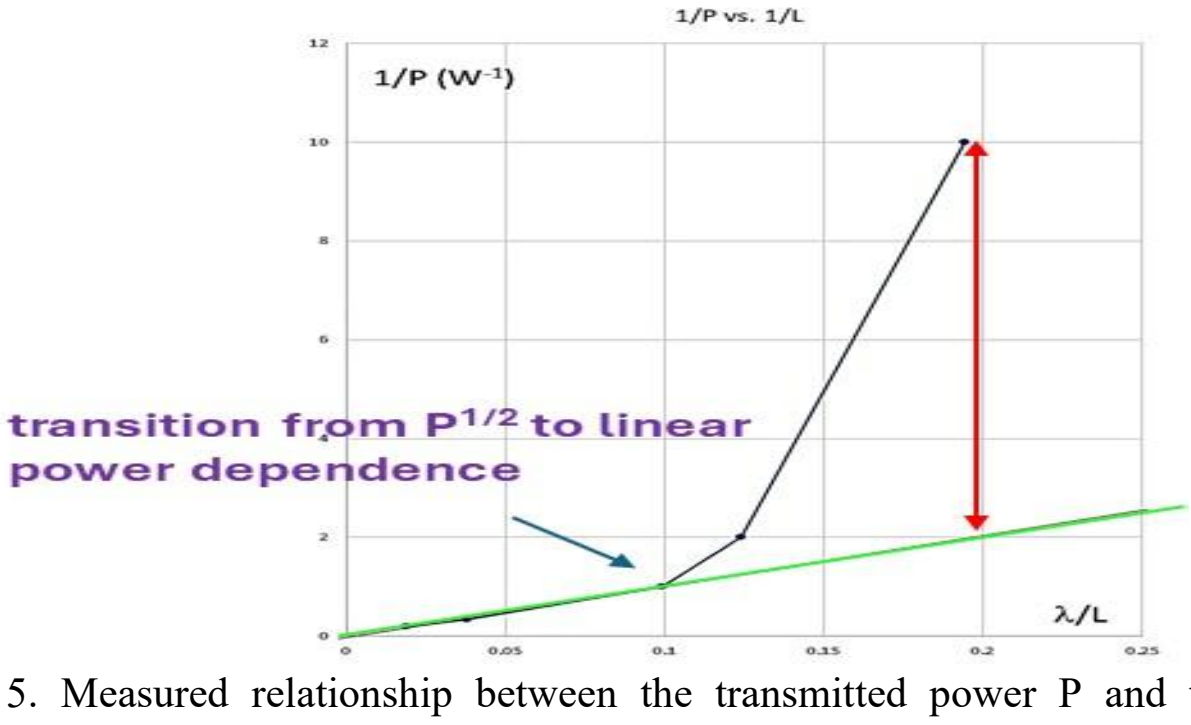

Fig. 5. Measured relationship between the transmitted power P and the communication distance L at 0.3 m antenna depth underwater shown as a functional dependence between 1/P and $\lambda_0$/L, The green line highlights P~1/L behavior at large distances. The red arrow demonstrates that near the antenna conventional RF fields dominate.

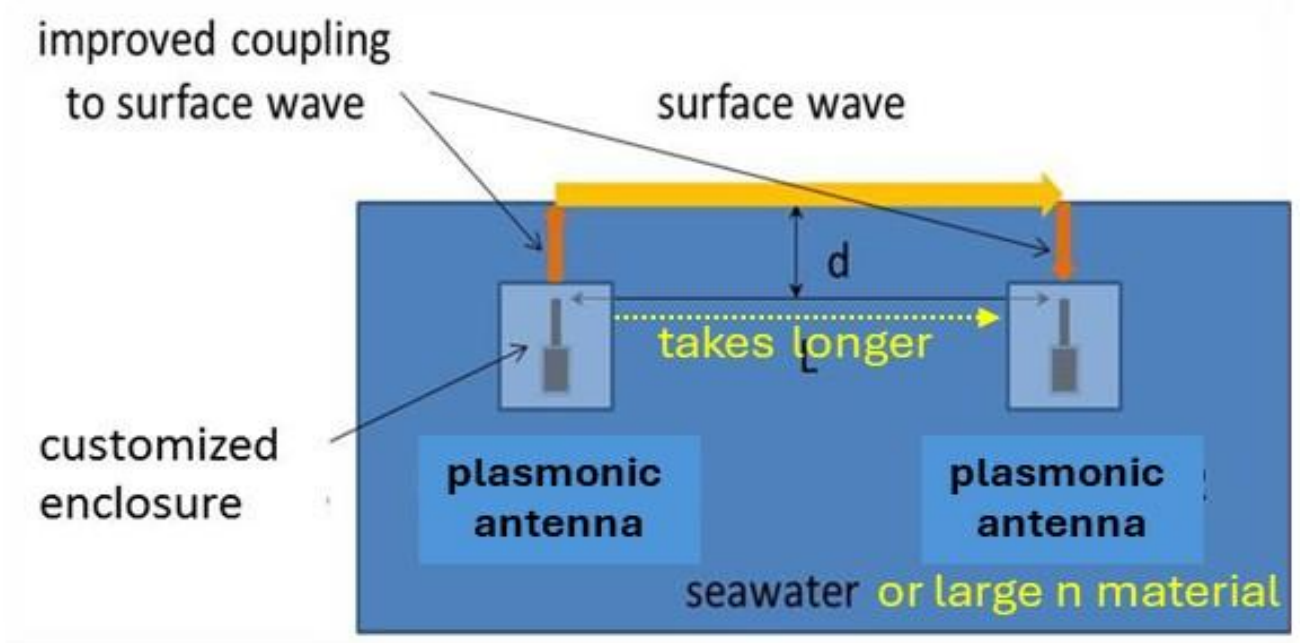

Fig. 6. Understanding of plasmonic antenna operation based on Fermat's least time principle.

## IV. Experiments Demonstrating Two-Dimensional Character of Signal Propagation Underwater

In reality, no plasmonic antenna is going to be perfect (for reasons given below), so that any such antenna or exciter will generate both conventional RF fields and surface electromagnetic waves. Therefore, it is very important to demonstrate actual excitation of radio frequency plasmonic modes in propagation experiments, and to measure the inverse proportionality between the received power and the communication distance, so that the 2D character of plasmonic signal propagation is proven. We have conducted such an experimental study with 30 MHz underwater plasmonic antennas in the case of the freshwater-air interface, as illustrated in Figs 4,5.

In order to facilitate data analysis, the plot in Fig.5 shows the measured relationship between the transmitted power P and the communication distance L at 0.3 m depth underwater as a functional dependence between 1/P and $\lambda_0$/L, where $\lambda_0$ is the free space wavelength. This way of data representation clearly reveals that at large distances from the antenna (see the data points near origin) P~1/L, while near the antenna the conventional P~1/$L^2$ behavior dominates, as indicated by the red arrow in the plot. Based on these measurements, we estimate that in this particular experiment the plasmonic antennas were about 45% efficient in a sense that 45% of emitted energy went into the surface wave channel, while the remaining portion went to conventional RF fields.

We should emphasize that while plasmonic antennas may potentially be made more efficient, conventional coupling is probably unavoidable due to such factors as surface scattering.

We should also note that at the most basic level the surface wave radio communication concept works based on Fermat's least time principle and Feynman's path integral formulation

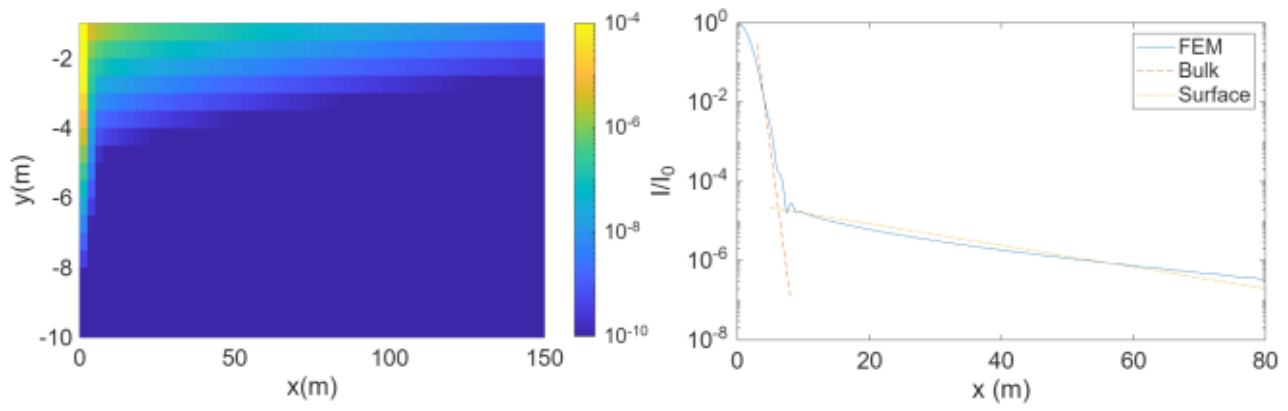

Fig. 7. (left) Intensity at the location of the detector (near water/air surface at x=0) as a function of the position of the emitter; (right) dependence of the intensity as a function of horizontal position of emitter, for emitter at $y = -1$ $(m)$; dashed lines illustrate decay length of volumetric and surface modes inside water and at smooth $(1/\beta = 10m)$ water/air interface (see Fig.2)

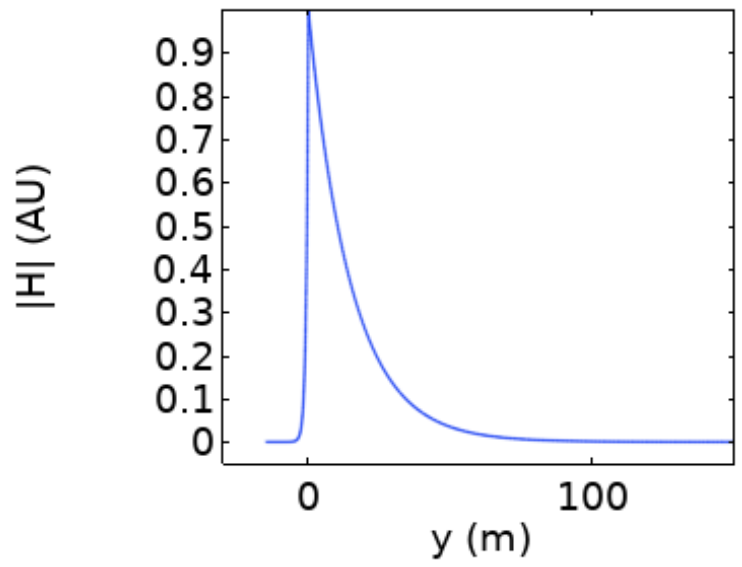

Fig. 8. Distribution of magnetic field in a surface guided mode in the water-ice-air interface geometry; permittivity of ice is $3.7 + 0.01i$; operating frequency is 30 MHz; effective mode index $n_g \simeq 1.0003 + 0.01i$ indicating a low loss non-radiative plasmon-like mode .

[17], and therefore it appears to be quite stable to such factors as sea state, etc. Indeed, as illustrated in Fig.6, Fermat's principle states that the path taken by a radio wave between two given points underwater is the path that can be traveled in the least time. Since the refractive index of water at RF frequencies is very large, the path between two underwater antennas which takes the least amount of time goes straight up to the water surface, continues through the air along the water surface, and straight down to the second antenna, as indicated in Fig.6. Moreover, Feynman's path integral formulation indicates that a radio wave propagates from point A to point B by taking all possible paths between these points [17]. Therefore, even if the water surface is quite wavy, there always be a suitable path for the radio waves in the air above the waves. These very basic arguments ensure reasonably stable operation of plasmonic underwater radio communication links. In the next section we will demonstrate that such links also operate quite well under ice.

To further analyze the behavior described above, we have numerically modeled underwater communication process, assuming a smooth water-air interface, with permittivity given by Eq.(3) with $1/\beta = 10\,m$. In these two-dimensional simulations, the (TM-polarized) wave was generated by a point (line) source, and the dependence of the electromagnetic fields as a function of the distance to the source and of the source depth has been analyzed.

The resulting dependence is shown in Fig.7. It is clearly seen that the communication distance between the emitter and the detector (located 1m below the surface) significantly increases when the source approaches the surface. Furthermore, the intensity decay along the water surface matches the decay rate of the surface plasmon polariton-like guided mode discussed above. This analysis further backs the surface-wave-based communication model described in Fig. 6.

## V. Plasmonic Communications Under Ice, Along Underwater Infrastructure and Along a Lake Floor

While a layer of ice may indeed perturb the underwater communication links, our theoretical modeling of this situation using COMSOL Multiphysics indicates that plasmonic modes do appear in the case of more complicated water-ice-air interfaces, and they may be used for efficient communication underwater under an ice layer. As illustrated by simulations depicted in Fig.8, the calculated mode index of the surface state in such more complicated geometries may also exceed the n=1 threshold, so that the surface mode appears to be a non-radiative low propagation loss plasmon-like state. These surface states were successfully used in our under-ice communication experiments depicted in Fig.9. These experiments also demonstrated our ability to directly communicate from about 7 m underwater through a 10 cm layer of ice to a radio located several meters above ice, without the need to break the ice surface. Moreover, the measured depth-distance performance under-ice appears to improve considerably compared to performance of similar plasmonic radios in the case of a simpler water-air interface [18]. These observations open up multiple additional use cases for radio frequency plasmonic communications.

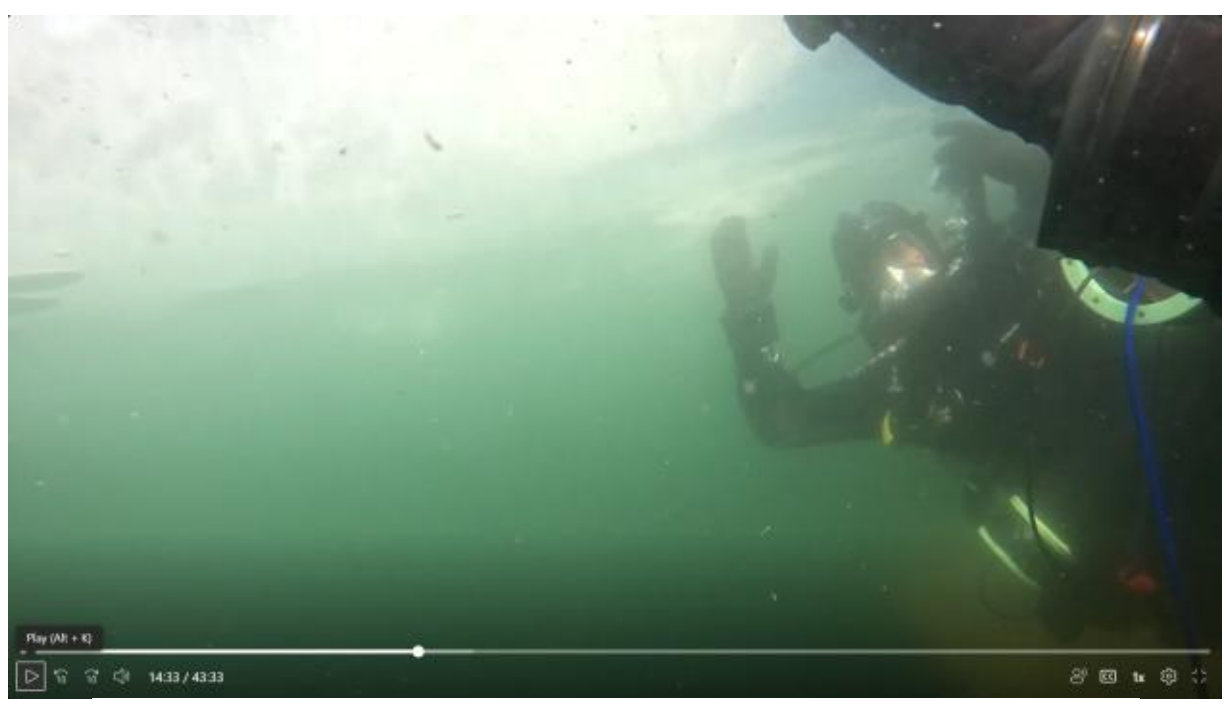

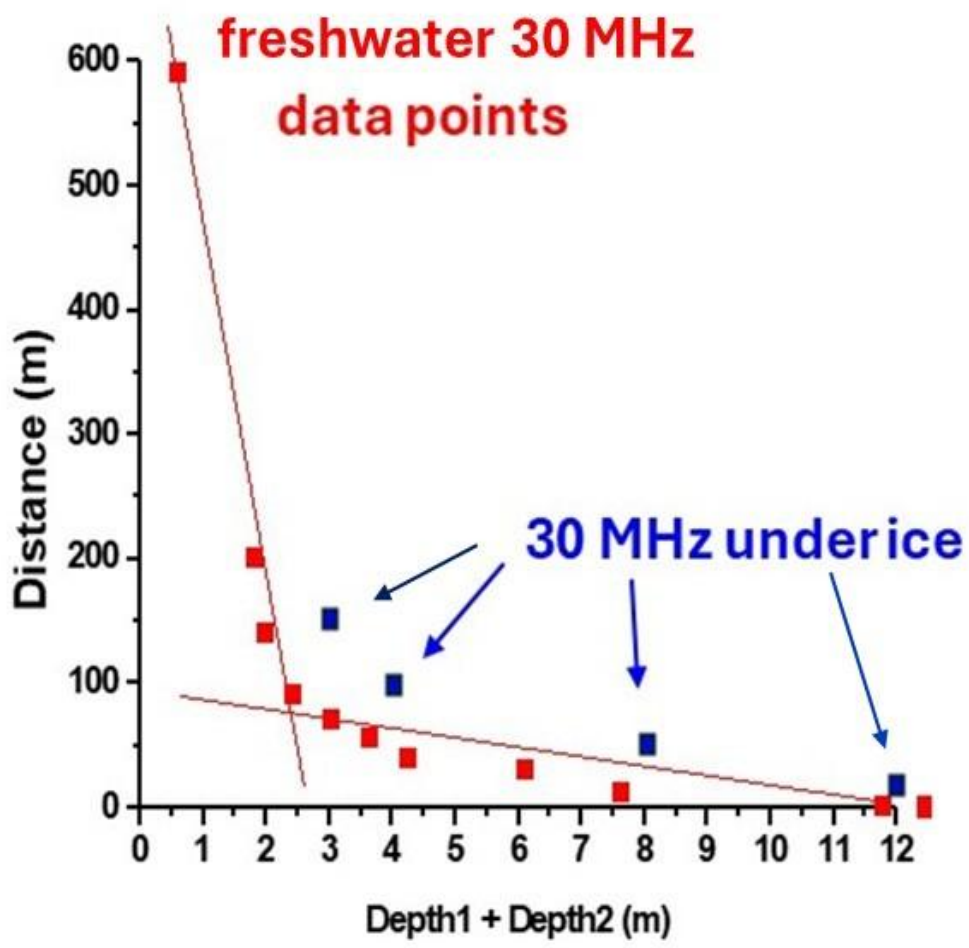

Fig. 9. Top panel: a diver operating a plasmonic radio under ice. Bottom panel: experimentally observed depth-distance performance under ice (blue data points) exceeds performance of a similar plasmonic radios underwater (red curves) as reported in [18].

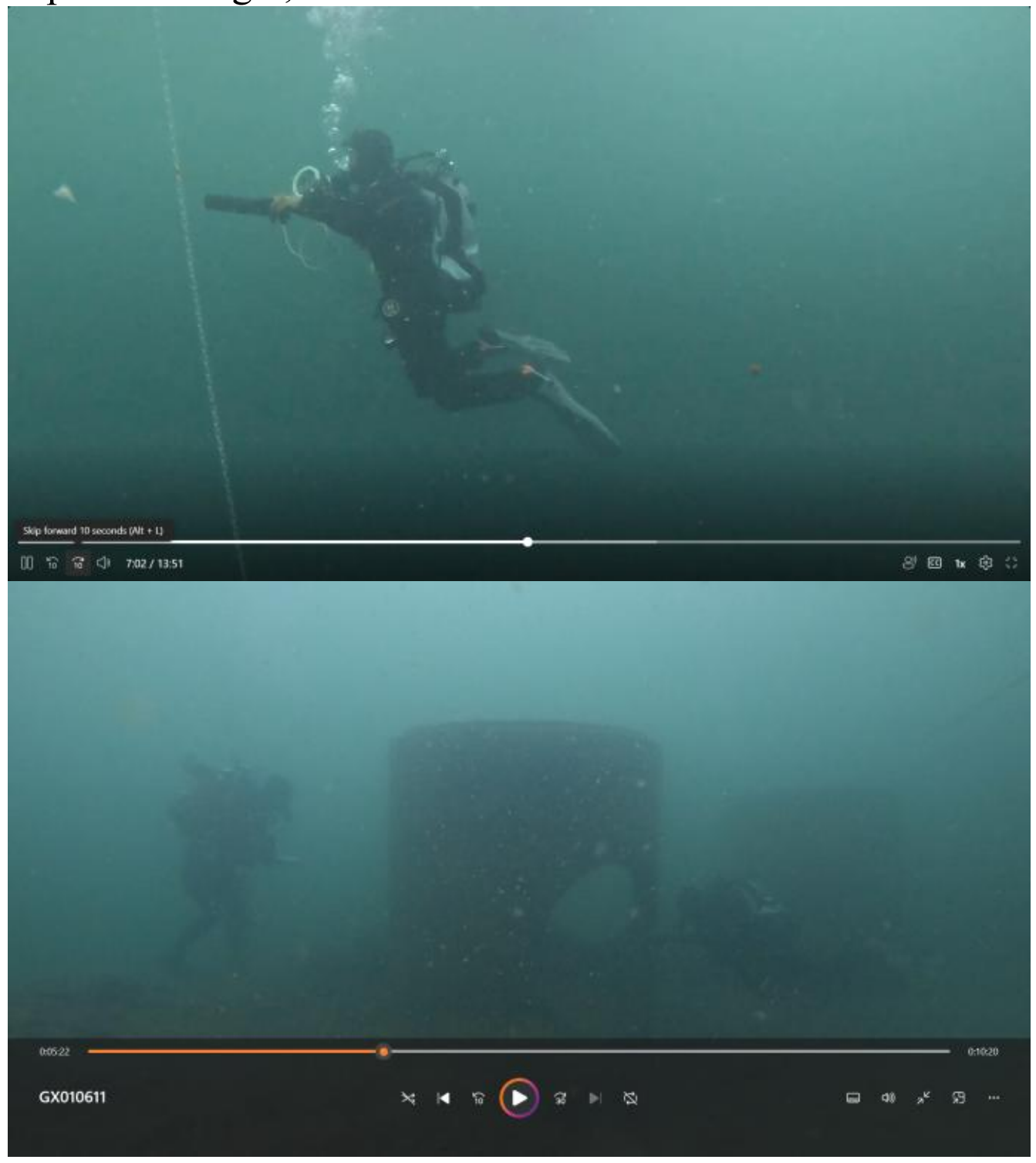

Fig. 10. Testing plasmonic communications along a metal chain suspended underwater (top panel), and along the lake floor (bottom panel).

We were also able to confirm theoretical expectations [5] that plasmonic communications may be successfully performed along an underwater infrastructure and along lake floor, as illustrated in Fig.10. In both situations the plasmonic modes exist due to drastic changes of conductivity across the interfaces involved [5]. The measured communication distance along the lake floor (located at about 18 m depth underwater) at 32 MHz was 15.3 m with 5 W power, which corresponds to approximately 15 wavelengths in fresh water at this frequency. However, the lakebed was quite rough, so that the measured communication distance was probably limited by surface scattering.

In the case of a metal chain, the measured propagation distance of radio signals along the metal chain in the same conditions reached a depth of about 12.2 m, which is consistent with our simulations of the effective mode index of a plasmonic mode propagating at 30 MHz along a 3-cm-radius metal cylinder in water – see Fig. 11a. Our measurements of radial distance from the chain (see Fig.11b) over which communication along the chain remains possible indeed confirm this mode existence.

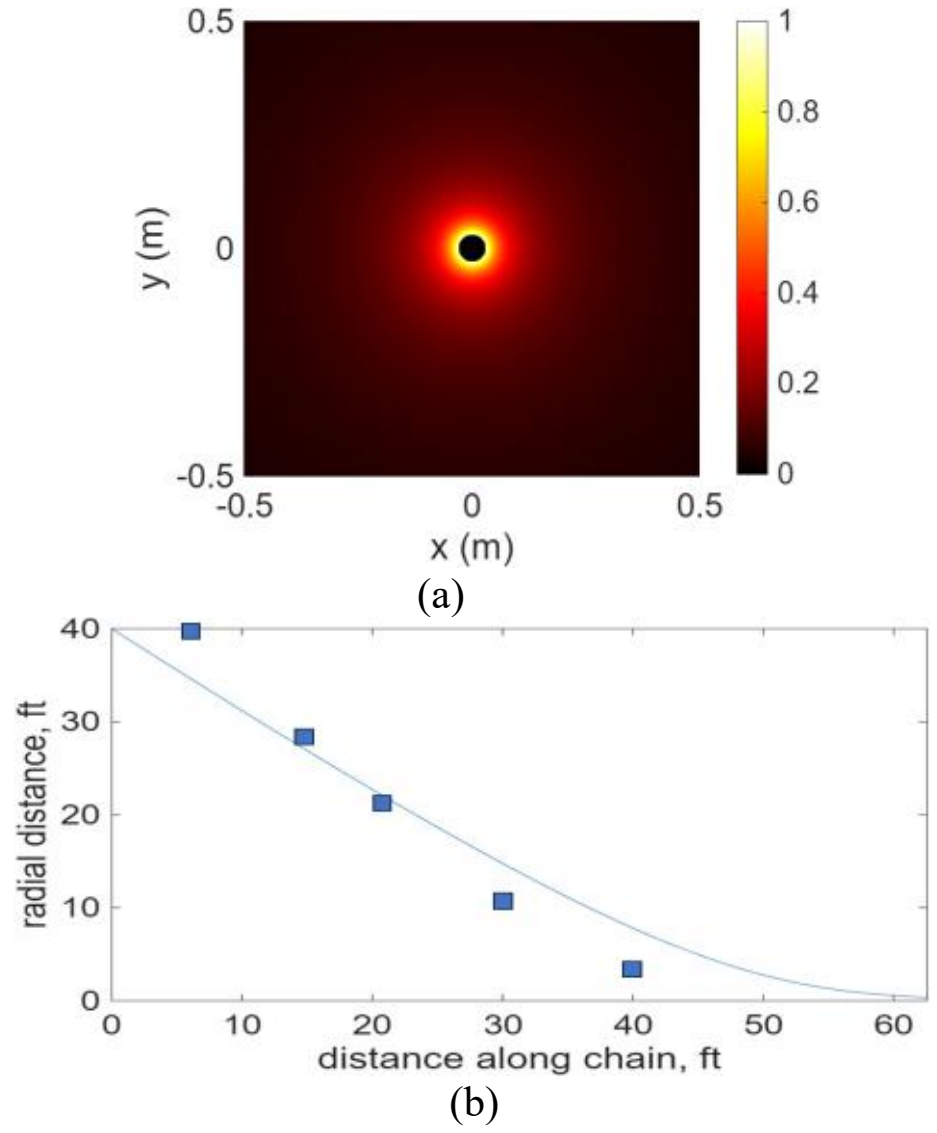

Fig. 11. (a) COMSOL Multiphysics simulations of a 30 MHz plasmonic mode propagating along the surface of a 3-cm-radius metal cylinder immersed in fresh water. (b) Measured radial distance r from the chain over which communication along the chain remains possible is plotted as a function of distance L along the chain. Measurements are compared with theoretical fit based on (a).

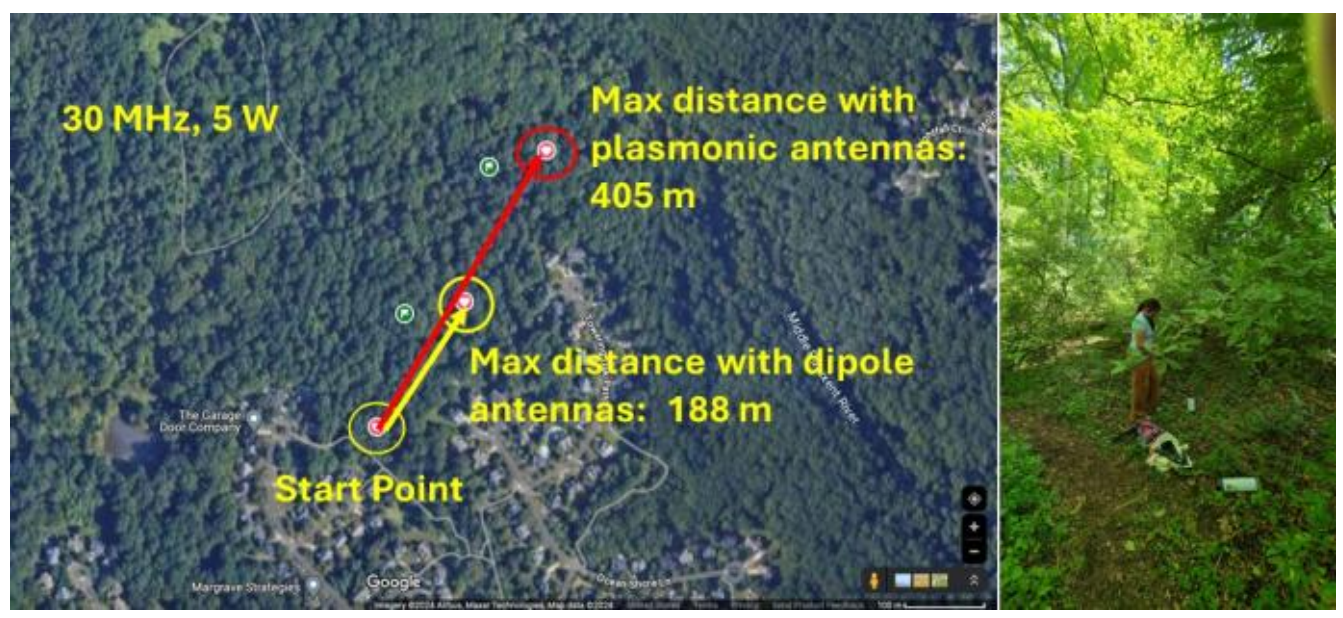

Fig. 12. Comparison of voice communication distances achieved using conventional native dipole antennas of the L3 Harris RF-7850M-HH radios and plasmonic 30 MHz antennas connected to the same radios in a dense forest area. Plasmonic antennas show more than 100% increase in communication distance. A photo of plasmonic communication experiment is shown on the right.

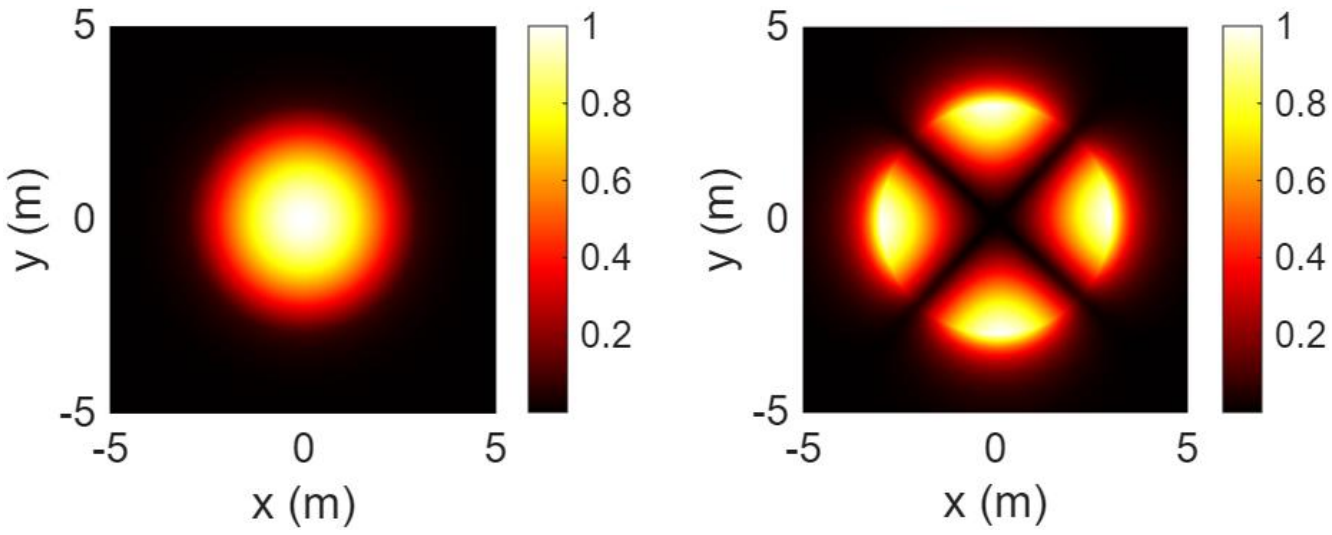

Fig. 13. COMSOL Multiphysics simulations of 30 MHz eigenmodes in a cylindrical underground tunnel; modal indices are 0.11+0.73i and 0.49+0.21i, respectively. The higher m mode localized near the ground-air interface exhibits 3.4x longer propagation distance.

## VI. Plasmonic communication in dense forests and underground

Our ability to use plasmonics for radio frequency communication in various underwater environments indicates that similar techniques should be applicable in many other “RF denied” situations. For example, it is well known that large amounts of moisture in vegetation prevent efficient propagation of conventional radio waves in the dense forests and jungle environments [19]. Since the dielectric properties of typical rocks and soil in the radio frequency range are similar to the properties of fresh water, it is also quite challenging to develop efficient radio communication systems underground [20]. Our experiments depicted in Figs. 12 and 14 indeed indicate that plasmonic communication may successfully supplement conventional radios in the jungle and underground environments.

First, we demonstrate that application of the effective potential formalism (see ref. [5] and Eq.(2)) looks especially straightforward in the case of a plasmonic mode propagating along the top of a jungle canopy. Based on the experimental results reported in [21], the dielectric properties of the canopy-air interface in the 10-1000 MHz frequency range are well described by the following approximation:

$$\varepsilon \approx 1 + i\varepsilon''(z) \approx 1 - i\alpha z \quad \text{at } -\xi < z < 0 \quad , \qquad (12)$$

where $\xi$ is the width of the transition layer between air and dense jungles, and $\alpha\xi$ is a small number ($\alpha\xi \sim 10^{-2} - 10^{-3}$, depending on the frequency range). The real part of the dielectric permittivity stays approximately constant across the interface, while the imaginary part changes in a linear fashion over a narrow transition region of width $\xi$ from $\varepsilon'' \approx 0$ in the air to some small constant number $i\alpha\xi$ deep inside the jungles. Application of Eq.(2) to such an interface results in appearance of an effective shallow potential well $V(z) = -3\alpha^2/4$ of thickness $\xi$, which guides a plasmon-like radio frequency surface electromagnetic mode along the top surface of the jungle canopy.

On the other hand, since the theoretical model of radio frequency plasmons on the water surface (see Section 2) is also applicable to the ground-air interface, this second plasmon-like mode may also be used in the jungles. Such additional surface-based radio frequency communication channels may become quite useful if the conventional bulk radio frequency transmission is either obscured or blocked by moisture or scattering by surface topography or vegetation. We have demonstrated that using plasmonic exciters we may indeed considerably extend radio communication range in a dense forest environment – see Fig.12. We should note that the same 30 MHz underwater exciter design was successfully used in this experiment.

While in the underground tunnel environments the delineation between plasmonic and “conventional” RF modes is not that obvious, our COMSOL Multiphysics simulations indicate that the zero angular momentum (m=0) modes and the higher angular momentum modes with fields more tightly localized near the ground-air boundary have substantially different propagation properties. An example of such simulations performed at 30 MHz is presented in Fig. 13, in which the higher m mode exhibits 3.4x longer propagation distance. Our experiments performed in the Luray (VA) caverns (see Fig.14) support these theoretical conclusions. As illustrated in this figure, using plasmonic antennas operating at 374 MHz, we were able to considerably extend communication range beyond a sharp corner, and we were able to communicate in the conventional RF “dead zone” located behind a large rock formation.

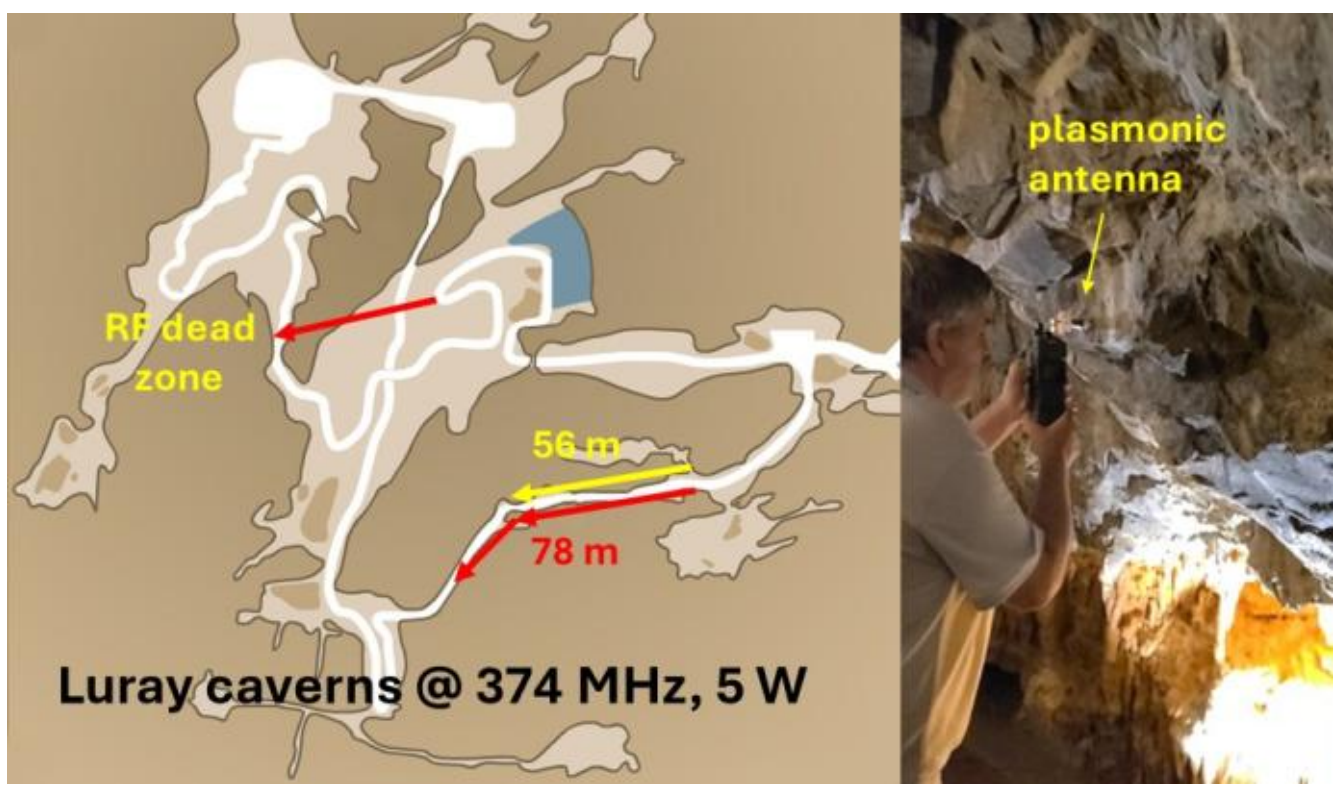

Fig. 14. An experiment with 374 MHz plasmonic antennas performed in Luray caverns. The plasmonic antenna performance is indicated by red arrows, as compared to conventional dipole antenna performance indicated by a yellow arrow. Plasmonic antennas extended communication distance by 40% beyond a sharp corner, and they were able to extend reliable voice communication into a conventional RF “dead zone” behind a large rock formation. A photo of plasmonic communication experiment is shown on the right.

Fig. 15. Compared to conventional GPRs schematically shown on the left, plasmonic radars penetrate much deeper underwater and underground, since the plasmonic signal scattering by an underground object directly affects the plasmonic field intensity on the surface, and the exponential hit on signal due to skin depth is only taken once.

## VII. Ground penetrating radars and underwater radars

The ability to communicate using plasmon-like surface electromagnetic waves underwater and underground indicate that plasmonics-based underwater radars and ground penetrating radars (GPR) may also be built successfully. Moreover, as illustrated in Fig.15, plasmonic radars should be able to penetrate considerably deeper underwater and underground, as compared to conventional GPR techniques. Since conventional GPRs rely on scattering of bulk RF modes by deep underground objects, the exponential decay of bulk RF waves (which is governed by conventional skin depth) affects the return signal twice on its way to and from the underground object – see the left panel of Fig.15. The reflected signal intensity measured by the conventional antenna may be estimated as

$$R \propto \frac{S}{r^2} e^{-4r/\delta} \quad , \qquad (13)$$

where S is the scattering cross section of an underground object, and δ is the bulk skin depth.

On the other hand, if surface electromagnetic modes are used, as illustrated in the right panel of Fig.15, the plasmonic signal scattering by an underground object directly affects the plasmonic field intensity on the surface, and the exponential hit on signal due to skin depth is only taken once:

$$R \propto \frac{S e^{-\frac{2|z|}{\delta'}}}{\rho} e^{-4\rho/\delta^*} \quad , \qquad (14)$$

where δ*>>δ is the surface wave propagation length, and z is the depth of the underground object. Thus, plasmonic radars should be able to penetrate much deeper underground and underwater, and unlike conventional GPRs, they are not blocked by underground water layers.

As illustrated by Fig. 16, we built such a prototype underwater plasmonic radar and demonstrated its basic functionality by detecting divers swimming approximately 1 m underwater (note that the ability of plasmonic GPRs to see behind a metal screen was demonstrated in our earlier paper [22]). We should also note that existence of plasmon-like surface electromagnetic waves at the water-air and ground-air interfaces strongly implies that superlens-like imaging of underwater and underground objects may be accomplished in a GPR configuration, which would lead to considerable increase in spatial resolution of the underwater and underground sensing and imaging techniques. The connection between superlensing and plasmonics in the visible range is discussed in [23]. A similar connection may be expected in the radio frequency range.

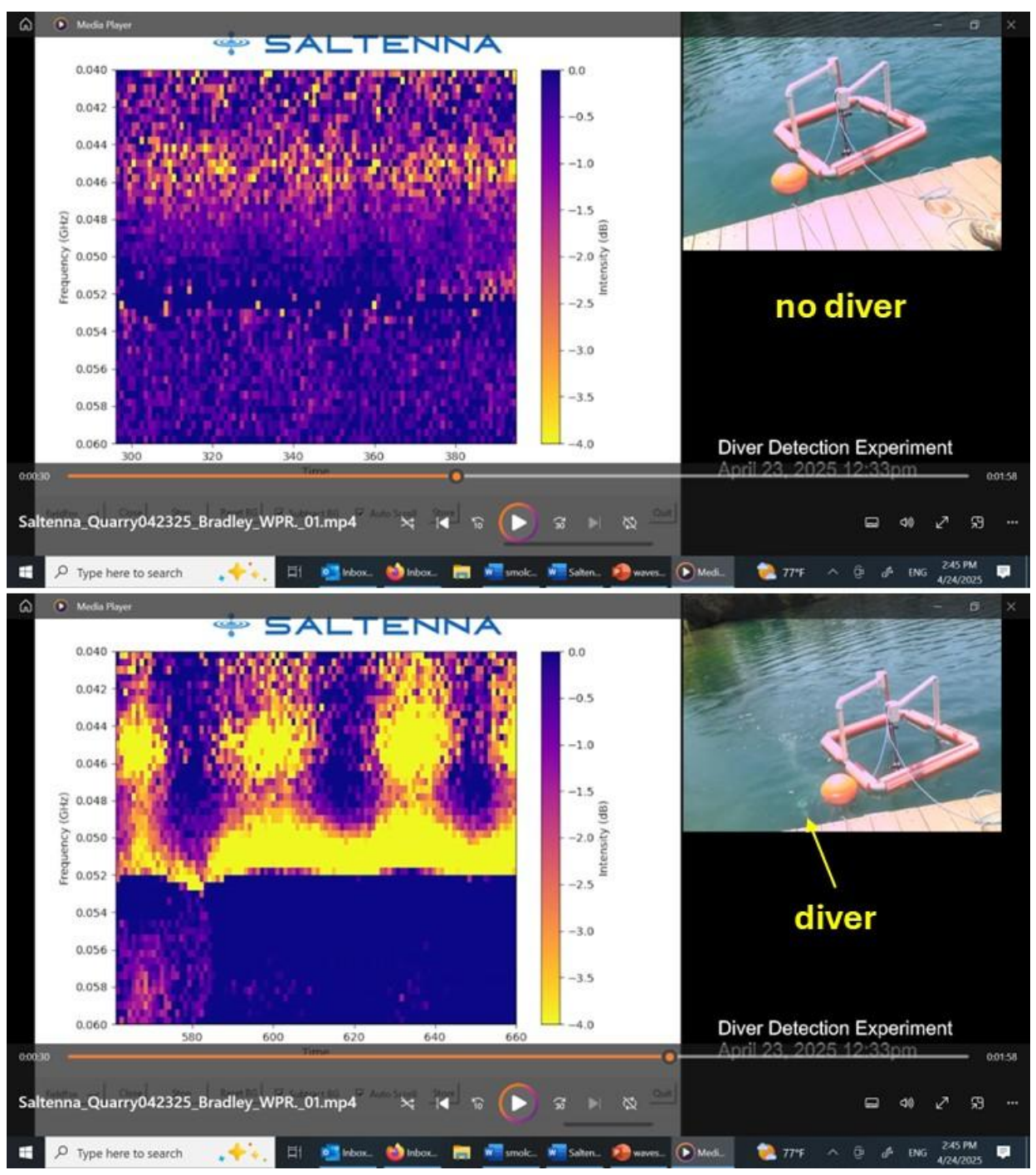

Fig. 16. Detection of a diver ~ 1 m underwater using a plasmonic radar (seen in the right panels) which is made of two 30 MHz plasmonic antennas mounted on the water surface. The temporal plots of S21 signal are shown on the left.

## VIII. Conclusions

In conclusion, we have demonstrated that low loss radio frequency plasmon-like surface electromagnetic waves propagate along rough interfaces between air and highly conductive media, such as water or metal. Unlike the well-known Zenneck wave solutions, these surface waves have mode index larger than 1, so that their propagation properties are generally similar to the properties of surface plasmons in the visible frequency range. Important applications of radio frequency plasmons were demonstrated, such as communication underwater, through ice, along metal infrastructure immersed in water, communication along a lake floor, and underwater plasmonic radar. Other applications of radio frequency plasmonic surface waves include communications underground and in dense jungles.